\documentclass[letterpaper, 10 pt, conference]{ieeeconf}  

\IEEEoverridecommandlockouts                              

\usepackage{graphicx}
\usepackage{tabularx}
\usepackage{cite}
\usepackage{algorithmic}
\usepackage{graphicx}
\usepackage{subfigure}
\usepackage{textcomp}
\usepackage{xcolor}
\usepackage{amssymb}
\usepackage{bbding}
\usepackage{multirow}
\usepackage{array}
\usepackage{amsmath}

\title{\LARGE \bf
Atrous Residual Interconnected Encoder to Attention Decoder Framework for Vertebrae Segmentation via 3D Volumetric CT Images
}

\author{Wenqiang Li$^{1}$, YM Tang$^{1,*}$, Ziyang Wang$^{2}$, KM Yu$^{1}$ and Sandy To$^{1}$
\thanks{$^{*}$YM Tang (e-mail: mfymtang@polyu.edu.hk) is the corresponding author.}
\thanks{$^{1}$Wenqiang Li, YM Tang, KM Yu and Sandy To are with the Department of Industrial and Systems Engineering, The Hong Kong Polytechnic University, Hung Hom, Hong Kong, China.}
\thanks{$^{2}$Ziyang Wang is with the Department of Computer Science, University of Oxford, Oxford OX1 3QD, the United Kingdom.}

}

\begin{document}

\maketitle
\thispagestyle{empty}
\pagestyle{empty}

\begin{abstract}


Automatic medical image segmentation based on Computed Tomography (CT) has been widely applied for computer-aided surgery as a prerequisite. With the development of deep learning technologies, deep convolutional neural networks (DCNNs) have shown robust performance in automated semantic segmentation of medical images. However, semantic segmentation algorithms based on DCNNs still meet the challenges of feature loss between encoder and decoder, multi-scale object, restricted field of view of filters, and lack of medical image data. This paper proposes a novel algorithm for automated vertebrae segmentation via 3D volumetric spine CT images. The proposed model is based on the structure of encoder to decoder, using layer normalization to optimize mini-batch training performance. To address the concern of the information loss between encoder and decoder, we designed an Atrous Residual Path to pass more features from encoder to decoder instead of an easy shortcut connection. The proposed model also applied the attention module in the decoder part to extract features from variant scales. The proposed model is evaluated on a publicly available dataset by a variety of metrics. The experimental results show that our model achieves competitive performance compared with other state-of-the-art medical semantic segmentation methods. 
\end{abstract}

\section{INTRODUCTION}


\begin{figure*}[htbp]
	\centering
	\includegraphics[scale=0.55]{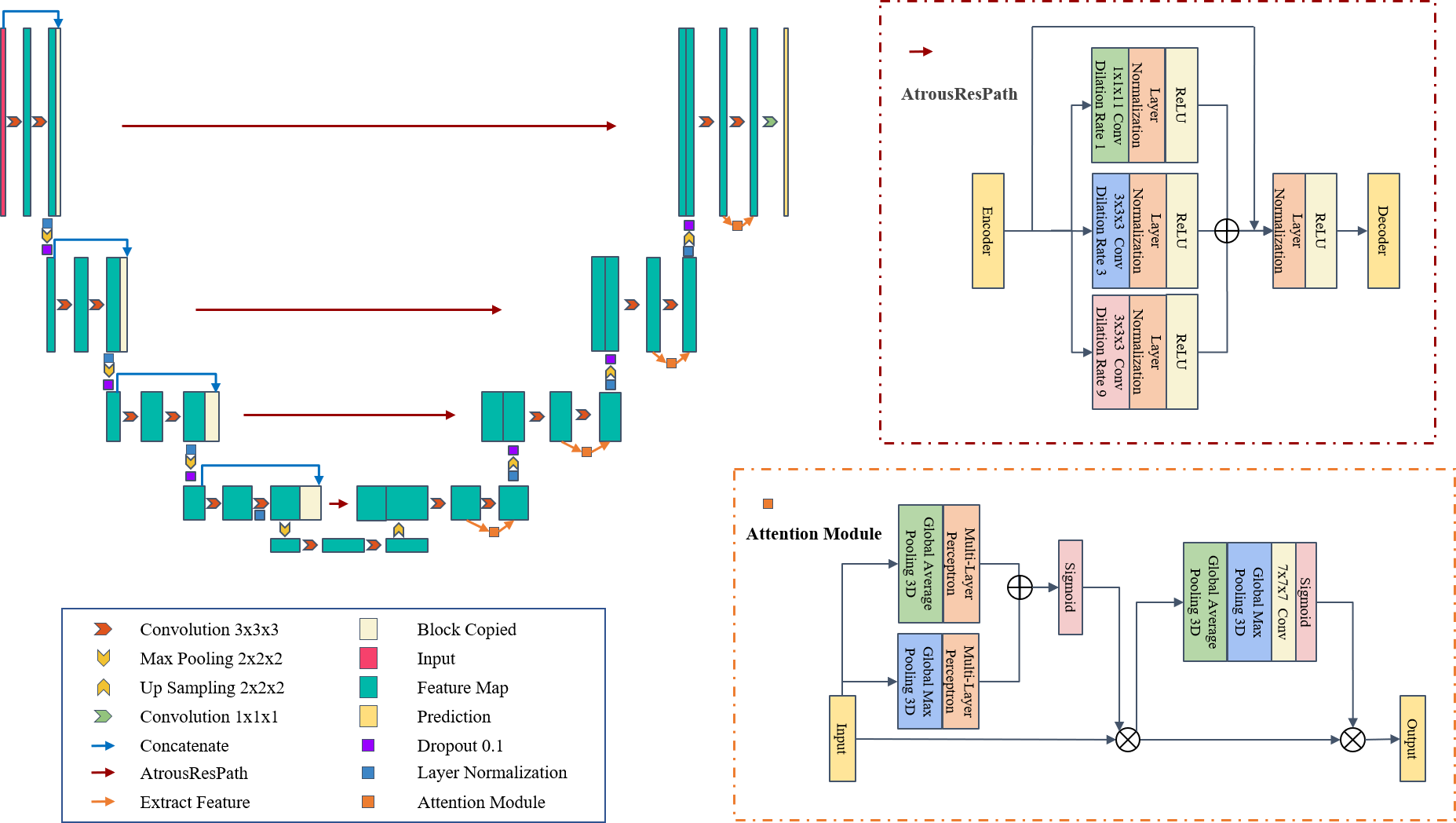}
	\caption{The Architecture of Atrous-ResUNet, Atrous Residual Path and 3D Attention Module.}
	\label{fig:framework}
\end{figure*}
Thoracic and lumbar vertebrae segmentation based on computed tomography (CT) is an essential step for many automated spine analyses, including computer-assisted surgery~\cite{knez2016computer}, detection of vertebral body fractures and spine deformities assessment~\cite{forsberg2013fully}. Spine represents not only the important central axis of musculoskeletal system, but also the flexible protective shell surrounding the spinal cord, the most important nerve pathway of human body. Computed Tomography (CT) image of spine is an essential tool for the evaluation of spinal lesions, which requires advanced computerized methods to support the diagnosis, treatment and intervention of physicians. Automated spine analysis requires not only the CT scans of spine, but also neck, chest, or abdomen that accidentally cover part of the spine. Therefore, the semantic segmentation algorithm of vertebrae needs to maintain robustness for a variety of image resolutions and different spinal coverage. 

A considerable amount of literature has been published on vertebrae segmentation. Recent studies for vertebrae segmentation trend to adopt deep learning technique, especially Deep Convolutional Neural Networks (DCNNs). UNet is developed by Ronneberger et al. in 2015~\cite{ronneberger2015u}, which is one of the most prominent semantic segmentation algorithms in biomedical area. 3D-UNet is proposed as an extension of UNet for 3D volumetric segmentation by replacing all 2D operations with their 3D counterparts~\cite{cciccek20163d}. While some research has been carried out on vertebrae segmentation based on 2D CT slices~\cite{wang2020rra}, there have been few empirical investigations into the vertebrae segmentation via 3D volumetric CT scans. The encoder-decoder based semantic segmentation algorithms are still having challenges in several ways. The information of feature maps decrease during the propagation from encoder to decoder. The reception field of filter is small, which cannot handle the situation of multi-scale object, as well as various image resolution. Lastly, the 3D volumetric data operation is memory-consuming, and the batch normalization has poor performance with mini-batch data. 

Based on the above considerations, we propose a novel 3D encoder-decoder framework, called Atrous-ResUNet. Firstly, Atrous Residual Path is designed to alleviate the spatial loss during the propagation from the encoder to the decoder. The design of Atrous Convolution can increase the reception field of filters. Secondly, to enhance the performance of the decoder, a 3D attention module based on the dimension of volume and channel is designed. Moreover, instead of batch normalization~\cite{ioffe2015batch}, we adopt layer normalization~\cite{ba2016layer} for faster and better convergence with mini-batch size under limited VRAM.
Subsequently, the evaluation was applied on the spine segmentation challenge of the CSI  2014~\cite{yao2012detection}. Compared with other state-of-the-art deep learning-based methods, the proposed model shows competitive performance with a Dice coefficient of 0.8928. 

The remaining part of this paper has been organized in the following way. Section II is concerned with the details of the proposed model. In Section III, we discuss the experimental setup and results. We conclude our study in Section IV.

\section{Methods}
The overall architecture of proposed Atrous-ResUNet is illustrated in Fig.~\ref{fig:framework}.
The main contributions of the proposed model comprise three components: Atrous Residual Path, 3D Attention Module in the decoder, and Layer Normalization. Firstly, we design Atrous residual Path based on Atrous Convolution~\cite{chen2017deeplab}, ResNet~\cite{he2016deep} and Inception~\cite{szegedy2016rethinking}. In the encoder, we designed a residual connection to alleviate the degradation problem. The residual encoder extracts the feature from the original CT images and the decoder with 3D attention module reconstructs the mask for corresponding CT images. 

\subsection{Atrous Residual Path}
Inspired by ResNet~\cite{he2016deep}, Inception~\cite{szegedy2016rethinking} and Deeplab~\cite{chen2017deeplab}, Atrous Residual Path is designed to reduce the spatial loss during the propagation from the encoder to the decoder, and alleviate the disparity between encoder and decoder. The architecture of Atrous Residual Path is shown in Fig.~\ref{fig:framework}, which comprises one 1x1x1 convolution filter with dilation rate = 1, two 3x3x3 convolution filters with dilation rate = 3 and 9, respectively. 
For each location i on the output y and a filter w, the general expression of Atrous Convolution is defined as follows:

\begin{equation}
y[i]=\sum_{k} x[i+r \cdot k] w[k]\label{atrous}
\end{equation}
where k denotes the kernel size, r represents the dilation rate and x is the input feature map. Besides, residual connect is also applied. Equation~\ref{resnet} define the general expression of residual connection.
\begin{equation}
y=F(x,{W_i})+x\label{resnet}
\end{equation}
where y is the output, and x denotes the input. $F(x,{W_i})$ is the function concatenated by 1x1x1 convolution filter and two atrous convolution filters. The operation $F(x,{W_i})+x$ is performed by a shortcut connection and element-wise addition. The output of the Atrous Residual Path from the encoder is concatenated with the corresponding up-sampling feature maps in the decoder.
\subsection{3D Attention Module}

Based on the spatial attention and channel attention~\cite{woo2018cbam}, we applied 3D channel attention and 3D volume attention module in the decoder to enhance the performance of decoder. As shown in Fig~\ref{fig:framework}, the proposed 3D attention module is integrate with two convolution layers in the decoder to enhance the essential information on feature map with Average-pooling layers, Max-pooling layers and Sigmoid functions. The attention module comprises two different attention is applied according to the dimension of channel and volume feature maps. The input of attention model is the feature map from previous convolution layer $\mathbf{F}\in R^{V\times H\times W\times C}$. $\mathbf{F}_{\text {avg}}^{\mathbf{c}},\mathbf{F}_{\text {max}}^{\mathbf{c}}\in R^{1\times 1\times 1\times C}$ are the feature maps generated by 3D average-pooling layer and 3D max-pooling layer on the dimension of channel. $\mathbf{F}_{\text {avg}}^{\mathbf{v}},\mathbf{F}_{\text {max}}^{\mathbf{v}}\in R^{V\times H\times W\times 1}$ are the feature maps generated by 3D average-pooling layer and 3D max-pooling layer on the dimension of volume. In the 3D channel attention, the shared Multi-Layer Perceptron (MLP) is applied to the features from pooling operations to generate the 3D channel attention map $\mathbf{M}_{\mathbf{c}} \in R^{1\times 1\times 1\times C}$. $\mathbf{W}_{0}$ and $\mathbf{W}_{1}$ are the weights of the shared MLP, where $\mathbf{W}_{0} \in R^{C/r\times C}$, $\mathbf{W}_{1} \in R^{C\times C/r}$ and r is the reduction ration, which is set as $8$. The output feature maps are merged by element-wise summation. The 3D channel attention and volume attention are illustrate in Equation~\ref{attention3} and Equation~\ref{attention4}. 
\begin{equation}
\begin{split}
\mathbf{M}_{\mathbf{c}}(\mathbf{F}) &=\sigma(\operatorname{MLP}(\text {AvgPool3D}(\mathbf{F}))+\operatorname{MLP}(\operatorname{MaxPool3D}(\mathbf{F}))) \\
&=\sigma\left(\mathbf{W}_{1}\left(\mathbf{W}_{0}\left(\mathbf{F}_{\mathbf{a v g}}^{\mathbf{c}}\right)\right)+\mathbf{W}_{1}\left(\mathbf{W}_{0}\left(\mathbf{F}_{\max }^{\mathbf{c}}\right)\right)\right)\label{attention3}
\end{split}
\end{equation}

\begin{equation}
\begin{split}
\mathbf{M}_{\mathbf{v}}(\mathbf{F}^{\prime}) &=\sigma\left(f^{7 \times 7\times 7}([\text { AvgPool3D }(\mathbf{F}^{\prime}); \operatorname{MaxPool3D}(\mathbf{F}^{\prime})])\right) \\
&=\sigma\left(f^{7 \times 7\times 7}\left(\left[\mathbf{F}_{\text {avg }}^{\mathbf{v}};\mathbf{F}_{\max }^{\mathbf{v}}\right]\right)\right)\label{attention4}
\end{split}
\end{equation}
where $\mathbf{M}_{\mathbf{v}}\in R^{V\times H\times W\times 1}$, $\mathbf{F}^{\prime}$ represents the generated feature map from the previous 3D channel attention, $f^{7 \times 7\times 7}$ is the 3D convolution operation with a kernel size of $7 \times 7\times 7$, $\sigma$ denotes the Sigmoid activation function, and $V, H, W, C$ denotes volume, height, width and channel, respectively. The overall expression of the 3D attention module is summarized as follows:
\begin{equation}
\mathbf{F}^{\prime}=\mathbf{M}_{\mathbf{c}}(\mathbf{F}) \odot \mathbf{F}\label{attention1}
\end{equation}
\begin{equation}
\mathbf{F}^{\prime \prime}=\mathbf{M}_{\mathbf{v}}\left(\mathbf{F}^{\prime}\right) \odot \mathbf{F}^{\prime}\label{attention2}
\end{equation}
where $\mathbf{F}$, $\mathbf{F}^{\prime}$, $\mathbf{F}^{\prime}{}^{\prime}$ denote the input, the feature map from the 3D channel attention and the final output, respectively, and $\odot $ represents element-wise multiplication.

\subsection{Layer Normalization}

Layer Normalization (LN)~\cite{ba2016layer} is applied in the proposed model to alleviate the issue of Batch Normalization (BN)~\cite{ioffe2015batch} with mini-batch. BN calculates the normalized statistics according to the number of samples. In this study, the batch size is 2, and the mean and variance of the four samples cannot reflect the global statistical distribution information. Due to limited hardware resources, BN is not applicable for the batch size larger than 2. If the batch size is too small, the calculated mean and variance are not enough to represent the entire data distribution. LN is an algorithm independent of batch size. LN no longer calculates the mean and variance of all the features in mini batch, but normalizes the three dimensions of VHW along the channel dimension, which overcomes the disadvantage that BN operation is sensitive to batch size. Therefore, to address the problem of BN under mini-batch, LN is applied. Layer Normalization is defined as follows:
\begin{equation}
\mu_i=\frac{1}{VHW}\sum x_i
\end{equation}
\begin{equation}
\sigma_i=\sqrt{\frac{1}{VHW}\sum (x_i-\mu_i)^2}
\end{equation}
\begin{equation}
LN(x_i) = \gamma\odot \frac{x_i-\mu_i}{\sqrt{\sigma_i^2+\epsilon}}+\beta
\end{equation}
where $\odot $ represents element-wise multiplication, $\epsilon$ is for numerical stability in case the denominator becomes zero by chance, $\gamma$ and $\beta$ are learn-able parameters.

\section{EXPERIMENTS AND RESULTS}



\subsection{Dataset} 
The spine and vertebrae dataset is acquired with Philips and Siemens multi-detector CT scanners by University of California~\cite{yao2012detection}. The dataset includes the thoracic and lumbar spine CT scans of 10 young adults. Each scan has up to 600 slices with 512$\times$512 resolution and 1mm inter-slice thickness. In this study, the original CT images are normalized and resized into the resolution of 256$\times$256 with the volume of 16 slices. Among 10 patients CT scans, 9 are used for training and the last patient's CT scan is used for validation and testing. In each scan, the thoracic and lumbar vertebrae are labeled as 1, and the rest are assigned with a value of 0. To enlarge the amount of training data and make them cover broader range, the training data are overlapped for every 5 slices. Moreover, various data augmentation mechanisms are also applied to augment the training data for both original CT slices and ground truth masks. The augmentation mechanisms include erosion, dilation, horizontal flip, vertical flip, random rotate 90 degrees, elastic transform, grid distortion and optical distortion, and each method is applied to $5\%$ training data, in a total of $40\%$ training data. 

\subsection{Experiment Setup}

The experiments are implemented under the environment of TensorFlow and Keras, and the hardware setup is Nvidia Tesla-V100 GPU with 16G Memory. For the training process of models, the batch size is defined as 2 due to the restriction of GPU memory. The learning rate and epoch are set to 0.00001 and 100 respectively for all the training processes, while Adam is the optimizer and ReLU is the activation function. The best model results are selected based on the value of validation loss. In the task of vertebrae segmentation, the ground truth only occupies a small area of the scan. In the predicted mask, the positive and negative pixels are strongly imbalanced. Therefore, instead of cross-entropy, we adopt Dice loss~\cite{milletari2016v} for optimization.



\subsection{Results and Discussion}
In this study, the dice coefficient is deployed as the major evaluation method. Besides, the standard performance matrix including accuracy, precision, recall, specificity is also used for evaluation and comparison~\cite{wang2020deep}. Dice coefficient is a set similarity measure function, which is usually used to calculate the similarity of two samples, and the value range is [0,1]. The segmentation performance evaluation method is defined as follows:
\begin{equation}
    Dice=\frac{2|M_{gt} \cap M_p|}{|M_{gt}|+|M_p|}\label{dice}
\end{equation}
where $M_{gt}$ represents the ground truth mask of vertebrae segmentation, and $M_p$ denotes the predicted mask of vertebrae segmentation.

\begin{figure}[htbp]
	\centering
	\includegraphics[scale=0.42]{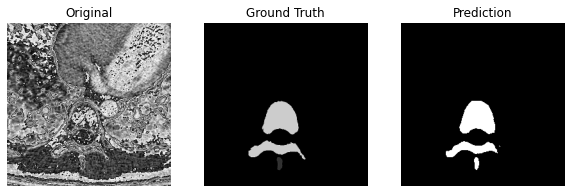}
	\caption{CT Slice Example, Ground Truth and Predicted Mask}
	\label{fig:example}
\end{figure}

The examples of the original CT slice, the ground truth segmentation mask of vertebrae and the predicted segmentation mask are shown in Fig~\ref{fig:example}, respectively. Besides, an ablation study is conducted to compare the performance of Batch Normalization and Layer Normalization in our model shown in Table~\ref{table:ablation}. 

\begin{table}[htbp]
	\caption{Semantic segmentation performance comparison among different algorithms on test set}
	\def\arraystretch{1.21}
	\begin{center}
		\begin{tabular}{p{2.6cm}<{\centering}|p{0.5cm}<{\centering}p{0.5cm}<{\centering}p{0.5cm}<{\centering}p{0.5cm}<{\centering}p{0.5cm}<{\centering}}
			\hline
			\textbf{Models} & \textbf{Dice}& \textbf{ACC}&\textbf{PRE}&\textbf{REC}&\textbf{SPE} \\
			\hline
			UNet-ResNet34 & 0.6114& 0.9749& 0.4424 & 0.9896 & 0.9746 \\
			UNet-VGG16 & 0.7377& 0.9770& 0.7237 & 0.7523 & 0.9871\\
			UNet-MobileNet & 0.7731& 0.9830& 0.6465 & 0.9614 & 0.9837\\
			UNet-Inceptionv3 & 0.7148& 0.9799& 0.5634 & 0.9775 & 0.9800\\
			UNet-DenseNet121 & 0.4773& 0.9693& 0.3144 & 0.9910 & 0.9689\\
			LinkNet-ResNet34 & 0.7468& 0.9803& 0.6514 & 0.8749 & 0.9839\\
			LinkNet-VGG16 & 0.8185& 0.9859& 0.7092 & 0.9678 & 0.9865\\
			LinkNet-MobileNet & 0.6776& 0.9777& 0.5240 & 0.9584 & 0.9782\\
			LinkNet-Inceptionv3 & 0.5267& 0.9711& 0.3604 & 0.9780 & 0.9710\\
			LinkNet-DenseNet121 & 0.5046& 0.9699& 0.3432 & 0.9620 & 0.9702\\
			FPN-ResNet34 & 0.7074& 0.9796& 0.5517 & 0.9855 & 0.9795\\
			FPN-VGG16 & 0.8191& 0.9861& 0.7011 & 0.9848 & 0.9862\\
			FPN-MobileNet & 0.6775& 0.9778& 0.5227 & 0.9623 & 0.9781\\
			FPN-Inceptionv3 & 0.7484& 0.9813& 0.6231 & 0.9369 & 0.9826\\
			FPN-DenseNet121 & 0.8081& 0.9836& 0.7718 & 0.8481 & 0.9893\\
			Residual-UNet & 0.8424& 0.9876& 0.7424 & 0.9735 & 0.9881\\
			3D-Residual-UNet & 0.8451& 0.9875& 0.7622 & 0.9481 & 0.9890\\
			Dense-UNet & 0.8680& 0.9894& 0.7795 & 0.9792 & 0.9898\\
			3D-Dense-UNet & 0.8745& 0.9894& 0.8206 & 0.9358 & 0.9916\\
			MultiResUnet & 0.8493& 0.9861& 0.8734 & 0.8265 & 0.9940\\
			\hline
			\textbf{Our Model} & \textbf{0.8928}& \textbf{0.9910}& \textbf{0.9910}& \textbf{0.9596}& \textbf{0.9923}  \\
			\hline
		\end{tabular}
		\label{table:performance}
	\end{center}
\end{table}


\begin{table}[t]
	\caption{Ablation study on Layer Normalization and Batch Normalization}
	\def\arraystretch{1.21}
	\begin{center}
		\begin{tabular}{p{2.6cm}<{\centering}|p{0.5cm}<{\centering}p{0.5cm}<{\centering}p{0.5cm}<{\centering}p{0.5cm}<{\centering}p{0.5cm}<{\centering}}
			\hline
			\textbf{Models} & \textbf{Dice}& \textbf{ACC}&\textbf{PRE}&\textbf{REC}&\textbf{SPE} \\
			\hline

			 Atrous-ResUNet with
			 Batch Normalization & 0.8375& 0.9873& 0.7329 & 0.9769 & 0.9876\\
			\hline
			 Atrous-ResUNet with
			 Layer Normalization & \textbf{0.8928}& \textbf{0.9910}& \textbf{0.9910}& \textbf{0.9596}& \textbf{0.9923}  \\
			\hline
		\end{tabular}
		\label{table:ablation}
	\end{center}
\end{table}
To comprehensively evaluate the proposed model's performance on the task of vertebrae semantic segmentation, we compare the results with various state of the art medical semantic segmentation algorithms, including pre-trained U-Net, LinkNet, FPN with ResNet34, VGG16, MobileNet, Inceptionv3 and DenseNet121 as backbone based on ImageNet, 3D U-Net, Dense-UNet, Residual-U-Net, 3D Dense-U-Net, 3D Residual-U-Net and MultiResUnet~\cite{ibtehaz2020multiresunet}. As shown in Fig.~\ref{table:performance}, our proposed model outperforms other state-of-the-art semantic segmentation methods in terms of Dice coefficient. As shown in Table~\ref{table:performance}, the proposed method achieves the best 0.8928, 0.9910, 0.9910, 0.9596 and 0.9923 in Dice, Accuracy, Precision, Recall and Specificity, respectively, outperforming other methods by a notable margin.

\section{CONCLUSION}
This paper proposes a novel framework to automated segment vertebrae via CT scans, which shows competitive performance compared with other state-of-the-art semantic segmentation models. Atrous Residual Path is designed to decrease the spatial information loss during the propagation process from the encoder to the decoder, while the Atrous convolution allows the model to have a larger reception field. Besides, Layer Normalization is applied for mini-batch 3D volumetric data. In the future, we will expand the research scope to Semi-Supervised learning with Atrous-ResUNet.

\addtolength{\textheight}{-12cm}  





\section*{ACKNOWLEDGMENT}

This research is supported by the Research Committee of The Hong Kong Polytechnic University through a studentship (project account code: RK3N).

\tiny
\bibliographystyle{ieeetran}
\bibliography{ref}

\end{document}